# Identification by Inelastic X-Ray scattering of bulk alteration of solid dynamics due to Liquid Wetting

M. Warburton[1], J. Ablett[2], J.-P. Rueff[2,4], P. Baroni[1], L. Paolasini[3], L. Noirez[1*]

[1] Laboratoire Léon Brillouin (CEA-CNRS), Université Paris-Saclay, CEA-Saclay, 91191 Gif-sur-Yvette Cédex, France
[2] Synchrotron SOLEIL, L'Orme des Merisiers, Départementale 128, 91190 Saint-Aubin, FRANCE
[3] ESRF, 71 Av. des Martyrs, 38000 Grenoble, FRANCE
[4] LCPMR, Sorbonne Université, CNRS, 75005 Paris, France
*corresponding author: laurence.noirez@cea.fr

**ABSTRACT:** We examine the influence at room temperature of the deposit of a water layer on the phonon dynamics of a solid. It is shown that the water wetting at the surface of an Alumina monocrystal has deep effects on acoustic phonons, propagating over several hundred µm distance and taking place on a relatively long time scale. The effect of the wetting at the boundary is two-fold: a hardening of both transverse and longitudinal acoustic phonons is observed as well as a relaxation of internal stresses. These acoustic phonon energy changes were observed by inelastic X-ray scattering up to 40 meV energy loss, allowing us to probe the solid at different depths from the surface.

**Highlight:** Using inelastic X-ray scattering, we show a two-fold bulk effect of hydrophilic wetting on solid phonon dynamics: depth-dependent hardening of acoustic phonons and internal stress relaxation. These effects take place over hours, with apparent different kinematics between hardening and stress relaxation. The results evidence a dynamic coupling between solids and liquids, echoing recent results on the identification of shear elasticity of mesoscopic liquids.

## 1. Introduction

The mechanisms governing the liquid solid interfacial energy are crucial for the understanding of coating, adhesion, decontamination, heat transfer or fluid transport. The surface energy determines the wetting processes that create an interfacial zone where the liquid energy becomes different from that of the bulk due to the imbalance between intermolecular interaction and surface attraction. While the interfacial effects have been widely studied and debated at the nanometer lengthscale [1], recent developments indicate that the liquid would be impacted even far from the interface [2]. This observation supports the picture of a correlated liquid state in agreement with the identification of mesoscopic liquid shear elasticity [3, 4]. The thermal gradient is interpreted in terms of an establishment of a local out-of-equilibrium liquid/solid state and thus points out the importance of considering liquids as a phononic propagating medium. In this frame, we attempt a new approach to answer the opposite question: does the wetting induce also a long range impact into the solid? In order to do so, the analysis of the dynamics of the monocrystal α-$Al_2O_3$ is examined at different depths in the solid relative to the wetted surface.

Our study focusses on the interaction of α-$Al_2O_3$ mono crystal and liquid water. α-$Al_2O_3$ has multiple industrial applications, from ceramics industries to catalysts [5]. This R3C crystal, generally known as corundum, has a rhombohedral symmetry [3, 4]. The α-Al$_2$O$_3$ c-plane (0001), also known as the basal plane, has been extensively studied [1,5, 8-18] due to complex interaction and surface changes that can take place when hydroxylated with relation to the various surface treatments [1]. In this study, we only consider the basal plane c-plane (0001). Three typical basal plane outer layers exist called-Al-I, Al-II and Al-III for single layer, double layer Aluminum and Oxygen terminated respectively [11]. We consider the case where the first atomic layer of the substrate is single layer Al terminated (Al-I). In this case, number of studies consider it as the most stable state at room temperature and atmospheric pressure [1, 17, 18]. The uppermost layer of α-Al$_2$O$_3$ hydroxylates in presence of water molecules, creating an intermediate Gibbsite-like layer between the substrate and water [10].

When the crystal has not been contaminated by the environment, the α-Al$_2$O$_3$ c-plane (0001) of the clean crystal displays strong hydrophilic properties, as exhibited by the low contact angle (θ <5°)[13]. This hydrophilic property is due to the creation of an intermediary Gibbsite-like layer at the interface [10, 18], after which an intermediary organized zone of liquid arises, that has properties closer to that of solid water. This intermediary zone results in strong polarization near the surface [19], and, therefore, a thermal gradient is expected [20, 21]. Despite the fact that polarization of water is predicted to be limited to the first few layers of liquid [19], recent experimental studies evidence non-Fourier heat transport at long range due to the interface [22], resulting in temperature gradients in the liquid several millimeters away from the solid surface indicating liquid vibrational changes. Reciprocally, it has been shown that the induction of interfacial temperatures affects the 'vibrational' surface states of the solid; i.e. at the nanometer scale [10]. However, to our knowledge, no study has been conducted to observe potential changes in vibrational states deep in the bulk of the solid (> 100nm depth [22]). The aim of this paper is therefore to shed light on potential dynamic changes in the crystal bulk due to wetting (down to 300μm), by means of an experimental analysis of the acoustic phonon variations using Inelastic X-Ray Scattering (IXS). The paper is organized as follows: the experimental methods, results and related discussion.

## 2. Experimental details

Inelastic X-Ray scattering was used to probe phonon dynamics of the solid substrate due to the negligible high energy photon beam of the Synchrotron radiation with the organic liquid layer compared to that of the electronically dense α-Al$_2$O$_3$ [23]. The experiments were carried out on beam line ID28 of the European Synchrotron Radiation Facility (ESRF). The incident beam energy was 17.794 keV, using the Si (999) reflection order for the high-energy resolution temperature-controlled monochromator and analyzer crystals, and providing an instrumental full width at half maximum energy resolution of 3 meV [23, 24].

The experiments were performed at room temperature and atmospheric pressure. Disk-like monocrystals of α-Al$_2$O$_3$ of 22mm diameter and 2mm thickness (Neyco manufacturer) of known orientation were used. These monocrystals were manufactured using the Kyropoulos method and were cut in the 2mm thick disk-like shape using a wire-saw technique. They were annealed by the manufacturer at 1500 °C to reduce distortion and stress after cut but, neither polished nor grinded further.

Prior to each series of inelastic measurements, the monocrystals were heated for 1h at 450°C in a temperature-controlled oven to anneal potential contamination of the sample, and relaxed back to 125°C at which they were kept before being then quickly placed vertically on the sample holder in a nitrogen-filled environment. The sample holder (in polymer resin) positions the sample without constraint.

Before each inelastic measurement set, the crystallographic orientation of the dry sample was precisely defined. Then, dry reference data sets were obtained by scanning at two points in the reciprocal space: ((0 0 12.5) and (-1 0 13.5)), in two different first Brillouin zones in the Γ-Z direction, corresponding to the reciprocal c-axis going through the center of the considered Brillouin zone, (0 0 12) and (-1 0 14) respectively. Each Brillouin zones correspond to incident beam penetration depths of roughly 150μm and 300μm respectively, for incident angles of 8.68° and 19.75° respectively. Each zone was scanned from -40 to 40 meV, corresponding to the energies in the acoustic range of phonons of α-$Al_2O_3$. Each measurement took roughly one hour and a half per scanned zone. The oriented monocrystal surface was then wetted by spraying room temperature distilled water with a syringe, closing back the chamber under continuous nitrogen flux, and then launching a long (twelve hours) acquisition to observe potential kinetic changes. Scans recorded during the first 90 min after wetting are named 'Wet 1h30', scans that took place between 90 and 180min 'Wet 3h' etc. As the samples were set vertically, no drop formation took place at the surface. A bottom hole was made in the sample holder to evacuate excess water. The dry and wet scans were alternatively looking at $Q$ = (0 0 12.5) and $Q$ = (-1 0 13.5), for the position of the main analyzer.

Further series of scans were carried out to observe in detail the kinematics of the acoustic phonon response of the wetted crystal. The energy window was reduced to 10 meV, resulting in faster scans and more data in shorter time intervals.

## 3. Results

Our experimental results indicate a series of notable effects to highlight.

*3.1 Hardening of acoustic phonons:*

Fig.1 illustrates the inelastic scattering displayed at a given orientation by the sample of α-$Al_2O_3$ (c-plane) as a function of the energy in the initial dry state and at different times after wetting. The inelastic curve shows an elastic-like central diffusion surrounded at higher and lower energies by two main inelastic scattering components, corresponding to longitudinal (L) and transverse (T) phonon scattering. The scattering shows also a central peak indicating a quasi-elastic scattering that is approximated by a Lorentzian curve while the inelastic contributions are modeled by Damped Harmonic Oscillator function (DHO) weighted by a Bose factor [25]:

$$S(Q,\omega) = I_c(Q)\frac{\Gamma_c(Q)}{\omega^2 + \Gamma_c(Q)^2} + [n(\omega) + 1]I(Q)\frac{\omega\Gamma(Q)^2\Omega(Q)}{[\Omega(Q)^2 - \omega^2]^2 + \Gamma(Q)^2\omega^2}$$

where $Q$ is the scattering vector, $\omega$ the frequency (or energy), $I_c(Q)$ and $I(Q)$ are the intensity of the quasi-elastic and of the inelastic peaks respectively, $\Gamma_C$ and $\Gamma$, the energy widths of the quasi-elastic peak and of the inelastic peaks respectively, $n(\omega)$ the Bose factor and $\Omega(Q)$ the frequency (or energy) of the inelastic peaks.

The comparison of the inelastic curve of the dry sample to the wetted ones indicates a notable energy shift of meV amplitude of the inelastic peaks towards higher values. The gains in energy of the acoustic phonons seem to indicate that water induced hardening takes place, i.e. higher energies of the inelastic peaks. Hardening occurs for both transverse and longitudinal acoustic phonons. The hardening effect is measurable down to roughly 150 μm depth.

The shift of the inelastic peaks evolves slowly and continuously towards higher energies. This energy gain can be also expressed in terms of shift of sound velocity. From the measurements carried out at different scattering vectors of the first Brillouin zone, the slope of the acoustic phonon frequency (or energy) as a function of the reduced scattering vector in the Γ–Z direction in the considered Brillouin zone provides the sound velocities of the material, both longitudinal and transverse, through the following relation [26], where $V$ is the sound velocity, and $f = \omega/2\pi$ the phonon frequency:

$$V = 2\pi \frac{df}{dq}$$

At roughly 150μm penetration depth, the speed of sound is increased by 3.86% and 15.87% for longitudinal (L) and transverse (T) acoustic phonon branches respectively (Table1), as a result of the hardening, as observed on Fig. 2. The present results are coherent with known values of α-$Al_2O_3$ [6, 7, 27, 28]. This hardening has an impact on rigidity tensor quantities through increase the speed of sound in the substrate [26].

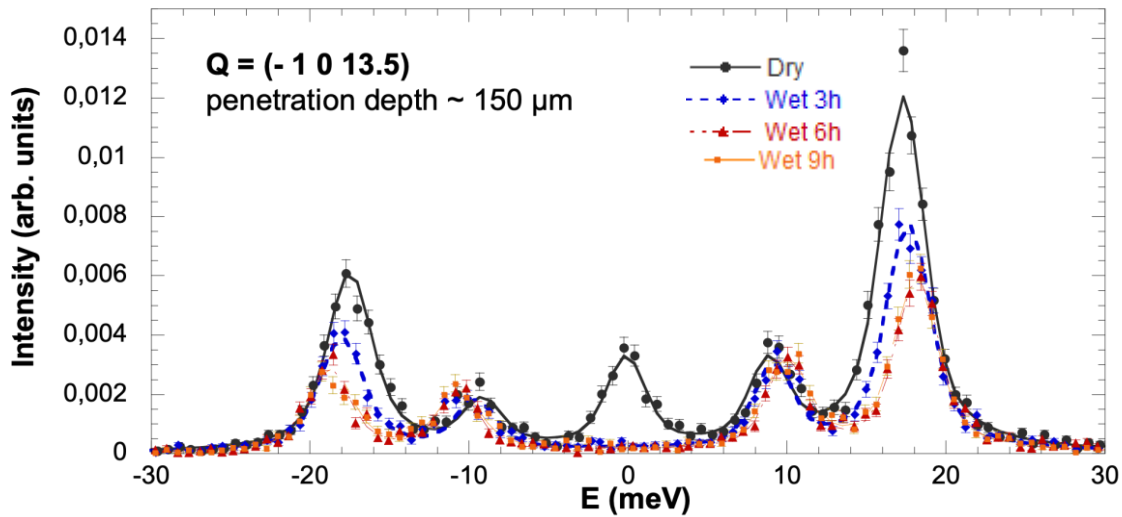

**Fig. 1**: Comparison of the inelastic scattering corresponding to the Dry α-Al2O3 (black), H2O wetted after 3hrs (blue), 6 hrs (red) and 9hrs (orange) at Q = (-1 0 13.5), corresponding to an incident beam penetration depth of roughly 150μm. A clear change of phonon spectra is observed for the acoustic phonon with inelastic peaks shifting to higher energies, corresponding to hardening. Measurements carried out on ID28 (ESRF) at room temperature and atmospheric pressure. The continuous line is the fit and the error bar in intensity is indicated. Transverse phonons are resolution limited (3 meV FWHM) while longitudinal phonons are not but do not show a clear trend with regards to wetting.

|  | Dry | 12hrs Wet 150 μm | 9hrs 30 Wet 300μm |
|---|---|---|---|
| $v_L$(km/s) | 10.95 | 11.37 | 11.05 |

| $v_T$ (km/s) | 6.13 | 7.10 | NA |

**Table 1:** Comparison of the sound velocities measured in the dry and wetted states of α-Al$_2$O$_3$ down to 150 and 300μm depth.

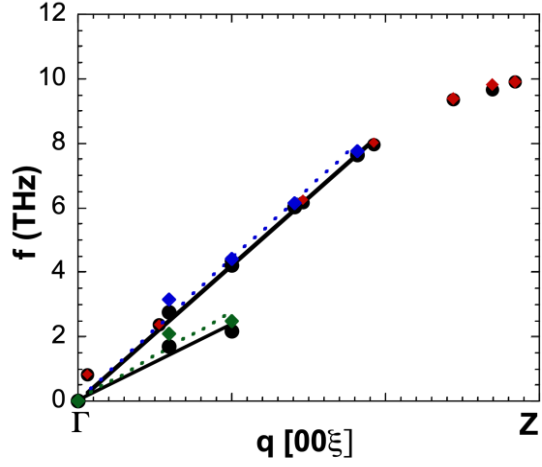

**Fig. 2**: Dispersion curve of dry (black full lines) and wet at 150 µm depth (dotted lines) in α-Al$_2$O$_3$ in $\Gamma$-Z direction. Blue points correspond to 12hrs wetted L phonon branch at 150 µm depth and green points to T phonon branch at 150 µm depth. Red lozenges correspond to longitudinal acoustic phonons at 300 µm depth. The energy at different scattering vectors $q$ was simultaneously recorded with the nine analyzers of the ID28 IXS beamline.

*3.2 Kinematics of hardening:*

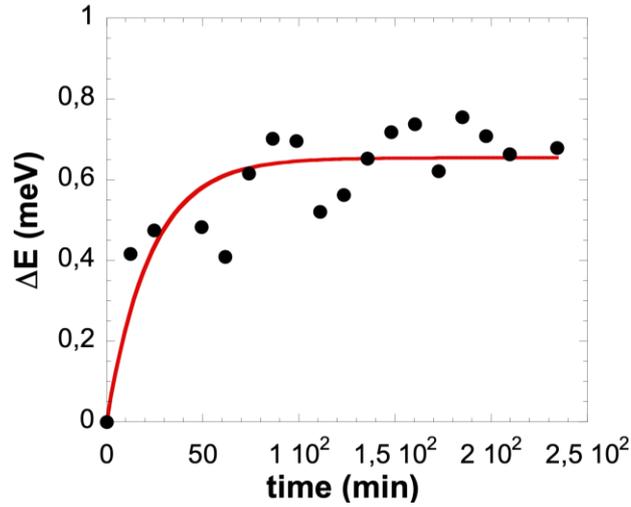

**Fig.3**: Kinematics of hardening at (-1.29 0 14.682), corresponding to 150 µm penetration depth. Experimental data is in black and in red is a relaxation fit of the form $\Delta E = \Delta E_{max}(1 - e^{-t/\tau})$. $\Delta E$ is the energy shift due to hardening, equivalent to a frequency shift (directly proportional) and t the relaxation time.

Fig.3 displays the kinetics of the shift of the inelastic peaks with respect to the dry state. It shows that the hardening remains relatively constant after 4hours. The longitudinal (L) acoustic phonon peak at (-1.29 0.14 14.682) shows a large time scale kinematic effect which might be fitted by a first order relaxation function, where $\Delta E$ is the energy shift with respect to the dry value and $\tau$ the relaxation time estimated at $\tau = 0.38$ hours. The long-term kinematics of hardening echoes the results from Fig.1, but, being at a different point in the Brillouin zone, the energy shifts cannot be directly compared. It also of interest to point out that the wetting induced energy shift remains stable for hours showing that the deep alteration of the interfacial vibrational dynamics is not affected by the energy of the incident radiation.

*3.3 Relaxation of internal stresses:*

Figs. 1 and 4 recorded at penetration depths of roughly 150 μm and 300 μm respectively shows the presence of a peak centered at $E = 0$ meV in the dry state. This quasi-elastic peak in the bulk structure is known to indicate internal stresses. The inelastic study shows that the surface wetting induces a strong relaxation of internal stresses as the intensity of the quasi-elastic peak decreases and then collapses (Fig.1 and Fig. 4). This effect takes place up to both 150 and 300 μm depth. Similarly to hardening, internal stress relaxation takes place on an hours long time scale, between 1h30 and 3hrs, as evidenced on fig 4.a and 4.b. The timescale of elastic stress relaxation is not the same as that of hardening.

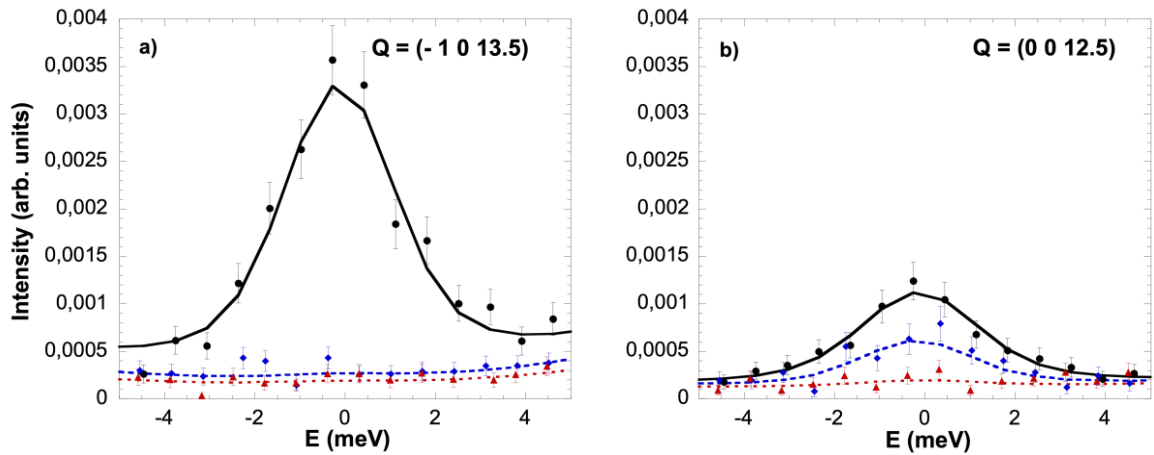

**Fig. 4** : Evolution of the quasi-elastic peak at a penetration depths of ) 150 μm b) 300 μm depth. The black circle corresponds to the data for the dry sample, blue is after 3h and 1h30 after wetting respectively, red is 6h and 4h30 after wetting respectively. Energy resolution: 3 meV (FWHM).

## 4. Discussion

The obtained results are new and also puzzling. While stress relaxation has already been observed [18] at the surface, to our knowledge, both hardening (Fig.1 and Fig.4) and deep internal stress relaxation are so far unreported effects of solid-liquid interfacial wetting down to these depths ( roughly 150 and 300 μm respectively). These wetting effects are characterized by long kinematics and macroscopic effects, such as the increase of transverse and longitudinal sound velocities of sizable value (see fig.2 and tab. 1). This pioneering study raises the question of the generic nature of this observation carried out on water wetting of alumina. From a fundamental point of view, recent advances on the identification of shear elastic properties of mesoscopic liquids may help elucidate our results. Previous experiments on the effects of the liquid-solid interface of hydrophilic materials (notably ceramics) indicate a highly correlated bulk effect in liquids leading to shear elasticity in water (and other liquids) at low frequency excitation and sub-millimeter scale [3,4], thus away from the theoretical high frequency model of the Frenkel theory [29], usually considered as the cornerstone of the molecular theory of liquids. Similarly, temperature changes without external thermal source at the interface has been observed, from thermal gradients of millimeters range in the quiescent liquid water [2] to mechanically induced temperatures when mesoscopic liquids are subject to shear-stress by a hydrophilic substrate [29-32]. These thermo-elastic effects, necessarily dependent on flow and fluid nature, are indicative of long-range correlations propagating far from the interface (>10μm). Therefore, as the temperature changes observed in the liquid are in energy windows of the same order of magnitude as the inelastic energy, it would make sense if such correlations were coupled to the solid dynamics and extend beyond the interface into the bulk of the solid, and strongly correlated phonon interaction would take place through the surface and the intermediate Gibbsite like layer and the organized water layer above. In agreement with the observation of the water temperature slightly decreasing in the liquid near the interface [2], the surface breaks the symmetry of energy transfer between liquid particles and wave-packets in water [32], a transport of energy from the liquid to the solid through H-bond optical phonon-like network of water has been proposed [33] and could explain the hardening reported in the current paper, echoing the phononic nature of liquids recently developed [34-38]. A direct observation of a static temperature profile in the liquid close to the solid surface suggests the establishment of a permanent metastable state that does not follow Fourier's law of heat conduction. An external stress would be capable of slightly shifting the equilibrium temperature into another thermodynamic (metastable) state in agreement with previous results reporting dynamic thermo-mechanical effects [30, 31]. Here, the dynamic imbalance at the interface is source of stress generating a metastable inter-changeably phonon state, and that is in both sides of the interface, length scale dependent. This imbalance is associated to extended characteristic lengths that might find origin in the cohesive nature of liquids [36] and further calls into question the Frenkel-Maxwell model already discussed [34-38].

Nonetheless, multiple questions still arise from our results. The presence of an elastic peak shows that the sapphire disks are in a preconstrainted state. The wetting acts on this preconstrainted state by relaxing it from stresses. The two different kinematics, seen in fig. 1 and 3 for hardening and fig.4 for the elastic peak, indicate that the first phenomenon to take place is that of internal stress relaxation, in less than 3hours compared to the 3 to 6 hours for hardening. As the internal stresses relax, a release of energy takes place and would lead to minimizing of acoustic phonon transfer throughout the material. This is supported by the increase of sound velocities (Fig.2). However, this picture is incomplete and raises a key question: up into which depth does elastic peak reduction and collapse takes place?

Another question relates to the phononic nature of liquids: Elton and Fernandez-Serra [33] describe optical phonons through the H-bond network, therefore at higher frequencies than that of the acoustic phonons in α-$Al_2O_3$. Therefore, what is the nature of the change of frequency (or energy) of said H-bond network optical phonons as they propagate through the surface?

## 5. Conclusion :

Using high resolution Inelastic X-Ray Scattering, we have highlighted a complex phonon interaction between liquid and solid, that leads to a change of acoustic phonon energy up until a given depth as well as internal stress relaxation in the solid and spans over hours. These results may suggest a new definition of the bulk frontiers for solids, as an extended intermediary zone dynamically affected by the presence of liquid, far beyond the usual bulk definition of solids (usually extended up to 100nm depth [20]). In essence, the above results show that the phononic nature of liquids, evidenced by their shear elasticity, has a strong impact on solid dynamics, and questions our understanding of the solid-liquid interface as a whole. Furthermore, the hardening of transverse phonon branches begs the questions: do liquids have transverse phonons and up to what length scale would they propagate? The present results shows that the inelastic scattering is a technique of choice to advance in the understanding of interfacial mechanisms.

**Acknowledgements:** This work has received funding from "Investissements d'Avenir" LabEx PALM (ANR-10-LABX-0039-PALM). We acknowledge the European Synchrotron Radiation Facility for provision of beam time on ID28. We would also like to thank Alexei Bossak for fruitful discussions.